\documentclass[amsmath,twocolumn,raggedbottom,nobalancelastpage,%
	aps,prb,10pt]{revtex4-1}
\usepackage{graphicx}
\usepackage{xspace}

\newcommand{\dvec}[1]{\ensuremath{\boldsymbol{#1}}}
\newcommand{\vk}{\dvec{\mathrm{k}}}
\newcommand{\vq}{\dvec{\mathrm{q}}}

\newcommand{\sgn}{\ensuremath{\mathrm{sgn}}}
\newcommand{\etal}{\textit{et al.}\xspace}
\newcommand{\dmudn}{\ensuremath{\frac{d\mu}{dn}}\xspace}
\newcommand{\mast}{\ensuremath{m^\ast}\xspace}
\newcommand{\muHF}{\ensuremath{\mu_{\mathrm{HF}}}\xspace}
\newcommand{\bise}{Bi$_2$Se$_3$\xspace}
\newcommand{\bite}{Bi$_2$Te$_3$\xspace}

\begin{document}

\author{D. S. L. Abergel}
\author{S. Das Sarma}
\affiliation{Condensed Matter Theory Center, University of Maryland,
College Park, Maryland 20742, USA}
\title{Two-dimensional compressibility of surface states on
three-dimensional topological insulators}

\begin{abstract}
We develop a theory for the compressibility of the surface states of
three-dimensional
topological insulators and propose that surface probes of the
compressibility via scanning single electron transistor microscopy will
be a straightforward way to access the topological states without
interference from the bulk states.
We describe the single-particle nature of the surface states taking into
account an accurate Hamiltonian for the bands and then include the
contribution from electron--electron interactions and discuss
the implications of the ultra-violet cutoff, including the universality
of the exchange contribution when expressed in dimensionless units.
We also compare the theory with experimentally obtained \dmudn as
extracted from angle-resolved photoemission spectroscopy measurements.
Finally, we point out that interaction-driven renormalization of the
Fermi velocity may be discernible via this technique.
\end{abstract}

\maketitle

%\section{Introduction and single particle analytical results}

Recently, there has been great interest in trying to engineer materials
with topologically protected states because of the applications that
such states may find in quantum computing.
In particular, it was predicted that certain three-dimensional (3D) 
layered crystals with
strong spin-orbit coupling and an inverted conduction band would have an
insulating band gap in the bulk, but topologically robust
conducting surface states with
linear low-energy dispersion protected by time-reversal symmetry.
\cite{zhang-h-natphys5}
Angle-resolved photoemission spectroscopy (ARPES) experiments on
crystals such as \bite and \bise have proven the existence
of these midgap surface states with nearly linear spectra. 
\cite{chen-science325, zhang-natphys6, xia-natphys5, zhu-prl107}
However, in most current samples, the Fermi energy is well above the bulk
gap, meaning that transport measurements of the surface states are
difficult because current can also flow in the bulk of the crystal,
\cite{steinberg-nl10} \textit{i.e.}, the bulk is a conductor rather than
an insulator even at low temperature.\cite{butch-prb81} This
can be mitigated to some extent by using thin films of the topological
insulator (TI) crystal\cite{kim-natphys8} or by doping the
crystal to compensate for the bulk conduction.\cite{ren-prb85} 
Therefore, it is essential that methods are found to both control and
characterize these topologically protected states.

In this Rapid Communication, we propose using single electron transistor
(SET) microscopy \cite{yoo-science276, yacoby-ssc111, martin-natphys4}
as a way of
mitigating the effect of the bulk states by approaching the surface
directly. 
Local capacitance measurements made via this technique can be converted
straightforwardly into the quantity \dmudn (where $\mu$ is the
chemical potential and $n$ is the carrier density in the surface state),
which is directly linked to the band structure and the electronic
compressibility. 
A great advantage of measuring this thermodynamic compressibility is
that it is a probe of the ground-state properties, which
incorporates single-particle many-body renormalization
directly, \cite{li-prb84} and thus quantitative information both about
the single-particle (SP) band structure and many-body renormalization
are obtained as a function of 2D carrier density (or,
equivalently, the 2D Fermi energy of the surface states).
We begin by calculating \dmudn for two
different commonly used approximations for the noninteracting band
structure of the Hamiltonian for the surface states of TIs such as \bise
and \bite and then take into account the electron-electron
interaction, demonstrating that this contribution is universal and
identical for both band structure approximations
when expressed in dimensionless units.
However, we show that the high dielectric constant of these
materials reduces the quantitative effect of the interactions at high
density.
We also argue that it may be possible to observe renormalization
of the quasiparticle Fermi velocity at sufficiently low density by this
technique.
We then compare the results of our analytical SP
calculations to estimations of \dmudn extracted from ARPES measurements
of the band structure. Finally, we comment on the role of disorder
and charge inhomogeneity on \dmudn in these systems.

If the surface corresponds to the (111) crystal direction
(\textit{i.e.}, parallel to the layered structure of the lattice), then
the simplest approximation for the band structure is the commonly used
linear form given by the Hamiltonian
\cite{zhang-h-natphys5} $H^l = \hbar v_F (k_y\sigma_x - k_x\sigma_y)$,
where $v_F$ is the band velocity, $\vk$ is the two-dimensional wave
vector in the plane of the surface, and $\sigma_{x,y}$ are Pauli
matrices in the real spin space. The spectrum of this Hamiltonian is
trivially the massless Dirac dispersion given by $\varepsilon^l_{\lambda
k} = \lambda \hbar v_F k$, where $\lambda=\pm 1$ denotes the band. 
The corresponding eigenvectors are two-component spinors in the spin
space. The 2D density of carriers in the SP limit is given
by $n = k_F^2/(4\pi)$, where $k_F$ is the Fermi wave vector and, as a
matter of convention, we say that zero density corresponds to the
chemical
potential being located at the Dirac point.
Elementary manipulations show that 
\begin{equation}
	\dmudn = \hbar v_F\sqrt{\frac{\pi}{|n|}}.
	\label{eq:lindmudn}
\end{equation}

\begin{figure}[tb]
	\includegraphics[]{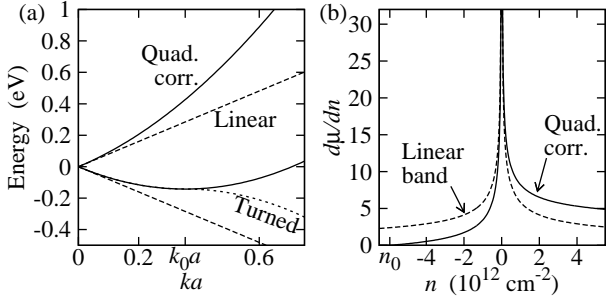}
	\caption{(a) Single-particle bands.
	The dotted line is the ``turned'' band structure where the 
	quadratic band is inflected at its turning point.
	(b) Single-particle \dmudn corresponding to Eqs. \eqref{eq:lindmudn}
	and \eqref{eq:quaddmudn}. 
	\label{fig:spbands}}
\end{figure}

A more sophisticated approximation for the band Hamiltonian including
the curvature in the valence band which is noticeable in the ARPES data
is  $H^q = \frac{\hbar^2 k^2}{2\mast} + \hbar v_F (k_y\sigma_x -
k_x\sigma_y)$, where \mast is the effective mass of the carriers
\cite{culcer-prb82} with spectrum $\varepsilon^q_{\lambda k} =
\frac{\hbar^2 k^2}{2\mast} + \lambda \hbar v_F k$.
These band structures are shown in Fig. \ref{fig:spbands}(a). In the
case with the quadratic correction, there is an unphysical turning point
in the valence band, which demands that some care must be taken when
describing this band. 
The turning point is located at $k_0 = v_F \mast / \hbar$,
the energy at this wave vector is $\varepsilon_0 = -\frac{v_F^2
\mast}{2}$, and the associated density is $n_0 = -{\mast}^2
v_F^2/(4\pi\hbar^2)$. 
Hence, defining wave vectors, energies, or densities which are greater
in magnitude than $k_0$, $\epsilon_0$, and $n_0$ requires a specific
definition of the band structure and this will become vital when the
interaction effects are incorporated.
We defer further discussion of this specific point until it is relevant.
However, for $n>n_0$, we use the relationship for the density above, and
find
\begin{equation}
	\dmudn = \frac{2\pi \sgn(n)\hbar^2}{\mast}
	+ \hbar v_F \sqrt{\frac{\pi}{|n|}}
	\label{eq:quaddmudn}
\end{equation}
where $\sgn(x)$ denotes the sign of the argument.
Figure \ref{fig:spbands}(b) shows the single-particle \dmudn for each
approximation as a function of density. 
Unless otherwise stated, we use $v_F = 5\times 10^5 \mathrm{ms}^{-1}$,
$\mast = 0.2m_e$ ($m_e$ is the electron mass), and $a=0.41\mathrm{nm}$
throughout this paper.
The quadratic correction preserves the dominant $1/\sqrt{n}$ behavior at
low density, but introduces an asymmetry between the conduction and
valence bands. 
In addition, \dmudn goes to zero at $n_0$. 

Note that some papers include a cubic term in the Hamiltonian, which
introduces a hexagonal distortion to the Fermi surface.\cite{fu-prl103, 
baum-prb85} However, our calculations show that this contribution makes
very little difference to the single-particle \dmudn in the range of
Fermi energy which we are interested in, and so we do not consider it
further.
%\section{Exchange contribution \label{sec:exchange}}

\begin{figure}
	\centering
	\includegraphics[]{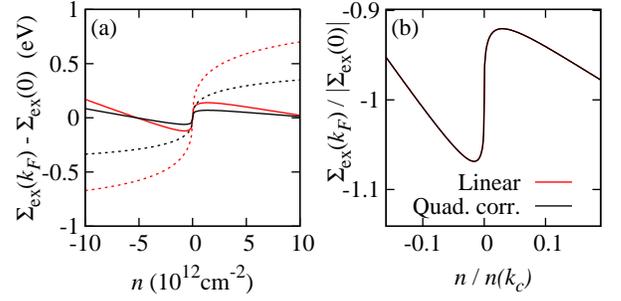}
	\caption{(a) The self-energy relative to its value at zero density.
	Solid lines indicate $k_c = 1/a$ and dashed lines show $k_c =
	2\pi/a$.
	(b) The self-energy normalized by its absolute value at zero
	density.
	Lines for both dispersions and all values of $k_c$ collapse
	onto each other.
	Throughout, the linear band is shown by red lines and the 
	quadratic band is shown by black lines. 
	\label{fig:sigma}}
\end{figure}
	
We now include the effect of the electron-electron interaction on the
compressibility.
Evaluating the self-energy contribution from the exchange interaction
gives
\begin{multline}
	\Sigma_{\mathrm{ex},\lambda}(\vk) = 
	-\frac{1}{8\pi^2} \int d^2\vq \sum_{\lambda'}
	V_C(\vq) f_{\lambda'}(\vk+\vq) \times \\ \times
	\left[ 1 + \lambda\lambda'\cos(\theta_{k+q} - \theta_{k}) \right]
	\label{eq:Sigma}
\end{multline}
where $V_C(\vq) = 2\pi e^2/(\kappa |\vq|)$ is the two-dimensional
Fourier-transformed unscreened Coulomb interaction, $\kappa$ is the
static dielectric constant of the environment,
$f_\lambda(\vk)$ is the occupancy of the state with wave vector $\vk$ in
band $\lambda$, and
$\theta_k$ is the angle that the wave vector $\vk$ makes with the $x$
axis. 
It is well known that for the compressibility, the exchange energy (the
Hartree-Fock term) is the dominant interaction correction, and this is
even more true for TI systems because of their very large background
dielectric constant. Therefore, we anticipate very small interaction
corrections beyond exchange in this problem.
Note that the only place that the band structure enters into this
expression is in the occupancy factors so that it is applicable to
systems described by both $H^l$ and $H^q$.
At zero temperature, the angular part of the integration of $\vq$ can be
computed analytically and yields
\begin{equation*}
	\Sigma_{\mathrm{ex},\lambda}(\vk) = 
	- \frac{e^2}{4\pi\kappa} \int_0^{k_c}
	dk' \sum_{\lambda'} Y_{\lambda k \lambda' k'} 
	\Theta\left( \mu - \varepsilon_{\lambda'k'}\right)
\end{equation*}
where the appropriate SP energy must be inserted into the
step function,
\begin{equation*}
	Y_{\lambda k \lambda' k'} = \frac{2}{k} \begin{cases}
	(k'+k) \left[ K(X) - E(X) \right] & \lambda\lambda' = 1 \\
	\frac{(k'-k)^2}{k'+k} K(X) - (k'+k)E(X) & \lambda\lambda'=-1
	\end{cases},
\end{equation*}
$K$ and $E$ are complete elliptic integrals of the first and second
kind, and $X = 2\sqrt{kk'}/(k+k')$.
The radial integral must be evaluated numerically. 
We note that the exchange integral of Eq.~\eqref{eq:Sigma} has an
ultraviolet high-momentum divergence arising from the linear dispersion,
which must be regularized through a high wave vector cutoff $k_c$. 
As is usual in condensed matter physics, there is a real cutoff in the
momentum arising from the lattice structure and, therefore, $k_c \sim
1/a$, implying that the interaction strength depends explicitly on the
short-distance lattice cutoff in the theory.
To illustrate the qualitative features of the physics and to determine
the dependence of \dmudn on $k_c$, we start by describing the $\kappa=1$
case in which interactions are the strongest.
This function is shown in Fig. \ref{fig:sigma}(a) for both the linear
band and the band with a quadratic correction for two different values
of the ultraviolet cutoff. 
Details of the definition of the band structure for the quadratic
correction are given below.
In Fig. \ref{fig:sigma}(b), the same data is shown but the units are
scaled to demonstrate how the self-energy depends on the cutoff $k_c$.
The linear and quadratic bands give identical results when the
self-energy is scaled by its value at zero density. 
We emphasize that this result shows the universality of the exchange
contribution, which is reminiscent of similar results found in
semiconductor heterostructures. \cite{dassarma-prb41, hwang-prb58}
The independence of the dimensionless many-body corrections of the
details of the TI band structure is an important new result of our work.

%\subsection{\dmudn for linear band}

\begin{figure}[tb]
	\centering
	\includegraphics[]{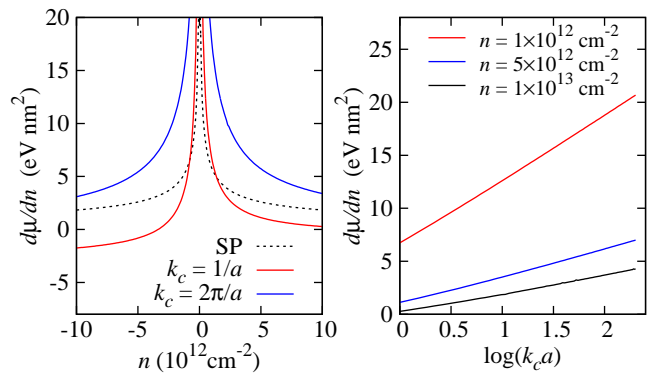}
	\caption{Linear band.
	(a) \dmudn as a function of density for the single-particle
	band, and two values of $k_c$. This illustrates that the choice
	of $k_c$ is crucial: small $k_c$ gives a negative
	contribution to \dmudn at higher densities, while large $k_c$ gives a
	positive contribution.
	(b) \dmudn as a function of $\log(k_c a)$ for different carrier
	densities.
	\label{fig:dmudn-linband}}
\end{figure}

The exchange self-energy is incorporated into the calculation for
$\dmudn$ in the following way. The Hartree-Fock chemical potential is
the sum of the SP kinetic energy and the exchange
self-energy: $\muHF = \mu + \Sigma_\mathrm{ex}$ and the corresponding
compressibility is given by $d\muHF/dn$, which may be computed
numerically.
When applied to the linear band structure associated with $H^l$, the 
expression in Eq.~\eqref{eq:Sigma} contains an ultraviolet divergence
and so the value of the high-frequency cutoff $k_c$ becomes important, as
hinted at in Fig. \ref{fig:sigma}.
In Fig.~\ref{fig:dmudn-linband}(a), we show \dmudn for the linear band
with the exchange contribution for two different physically reasonable 
values of $k_c$.
In all cases, the $n^{-\frac{1}{2}}$ behavior persists at low density.
For $k_c= 1/a$, the exchange causes a reduction in \dmudn relative to
the SP case for all but the lowest densities and may become
negative for large valence band doping.
In contrast, $k_c=2\pi/a$ gives an enhancement to \dmudn.
This indicates that the precise value of the ultraviolet $k_c$ has an
important quantitative effect on \dmudn.
In Fig. \ref{fig:dmudn-linband}(b), we show the dependence of \dmudn on
$k_c$ for three different values of the density. Noting the logarithmic
scale on the $k_c$ axis, we see that $\dmudn \propto \log( k_c a)$.
Hence, in order to make quantitative predictions of the behavior of
\dmudn at finite density, some physical intuition must be used for
choosing the ``correct'' value of $k_c$.
It has also been shown that \cite{baum-prb85} the cubic term, which we do
not consider, can introduce a scale above which the cutoff dependence is
negligible for the calculation of magnetization. In our case, the
integrand remains finite at all $k$ even with the cubic term, but
the value of $k_c$ extracted from this reasoning could be imposed on
the self-energy calculation.
The logarithmic dependence of the interaction effect on $k_c a$
translates into a slow logarithmic renormalization of the coupling
strength in units of $k_c/k_F$ which could manifest itself at very low
carrier densities.
\footnote{Note that in graphene, the cut-off may be set by
stipulating that number of states in each band corresponds to one
electron per atom but no such reasoning exists for TI crystals.}

%\subsection{\dmudn for quadratic correction \label{sec:turned}}

\begin{figure}[tb]
	\centering
	\includegraphics[]{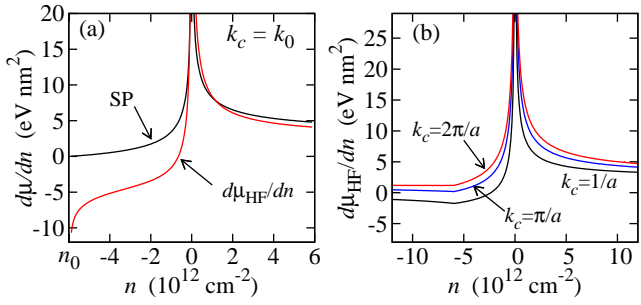}
	\caption{The quadratic correction. (a) \dmudn for $k_c = k_0$.
	(b) \dmudn for various values of $k_c>k_0$. 
	\label{fig:dmudn-quad}}
\end{figure}

For the quadratic band, Eq. \eqref{eq:Sigma} still holds, but the
appropriate dispersion must be used in the Fermi function. The shape of
the valence band introduces a further complication. Taken at face value,
the band minimum at $k_0$ should be interpreted as producing a Fermi
surface with nontrivial topology, and contributions to \dmudn should be
included from both branches of the valence band. However, this does not
correspond to the experimental data (see, for example, Ref.
\onlinecite{hasan-rmp82}) where the turning point is not seen.
The first (and most intuitive choice) is to set $k_c=k_0$. 
This may be physically reasonable because the topological surface bands
typically merge with the bulk bands at roughly this wave vector.
Another possibility is to artificially turn the band structure for $|k|
> k_0$ in the valence band [as illustrated by the dotted line in Fig.
\ref{fig:spbands}(a)] so that for $|k|>k_0$ we have $\varepsilon^q_{-k} =
-\frac{\hbar^2 k^2}{2\mast} - \hbar v_F k$. 
Then, from a technical point of view, $k_c$ can be set arbitrarily high
and the quadratically dispersing nature of the band is retained.
Figure \ref{fig:dmudn-quad}(a) shows the first of these cases. 
Since $k_c$ is relatively small, the density can reach the regime where
the self-energy is decreasing rapidly, $n/n(k_c) \sim 0.1$. 
In that case, the negative slope of the exchange energy is stronger than
the positive slope of the SP chemical potential so that
$\mu_{\mathrm{HFA}}$ is a decreasing function of $n$ for $n<0$, implying
that $d\mu_{\mathrm{HFA}}/dn$ is negative.
We believe that these questions about the precise
value of the ultraviolet cutoff in this problem can actually be
answered through a careful comparison between experiment and theory on
the TI compressibility.

For the turned band structure with $k_c > k_0$, the evolution of
$\Sigma_\mathrm{ex}$ with increasing $k_c$ means that the sign of the
exchange contribution can be positive or negative.
For a given density, we have $\dmudn \propto \log(k_c a)$, as was the case
for the linear band.

%\subsection{Effect of dielectric environment}

\begin{figure}[tb]
	\includegraphics[]{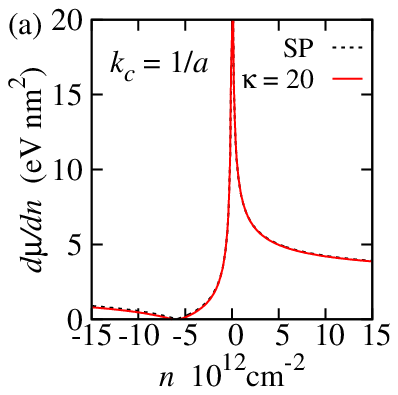}
	\includegraphics[]{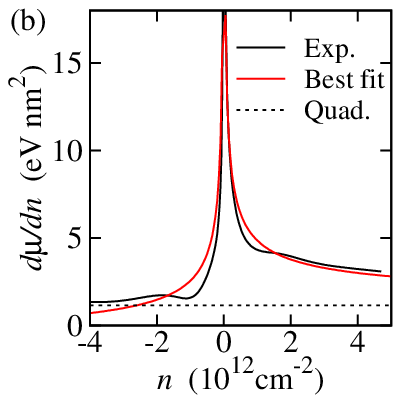}
	\caption{(a) The effect of electron-electron interactions for
	the quadratic band with $\kappa=20$ and $k_c = 1/a$.
	(b) Black line: \dmudn computed from band structure 
	extracted from ARPES data in Zhu \etal \cite{zhu-prl107}
	(see text for details). Red line: Least-squares best fit to the
	analytical result of Eq. \eqref{eq:quaddmudn}. The parameters
	extracted from the conduction band are $\mast = 0.417m_e$ and $v_F =
	1.00\times 10^6 \mathrm{ms}^{-1}$.
	\label{fig:kappa-exp}}
\end{figure}

The static dielectric constant of crystals such as \bise and \bite is
estimated to be \cite{richter-pssb84, butch-prb81} greater than 50,
reducing the strength of the interactions.
We assume that the effective dielectric constant is the average of that
in the material and the vacuum, and take a physically reasonable value of
$\kappa=20$. Figure \ref{fig:kappa-exp} shows that the exchange
contribution to \dmudn is reduced so much that it is essentially
nonexistent. Hence, for the purposes of examining the compressibility
of the surface states, the SP calculation is likely to be quantitatively
sufficient.

At this point, we pause to comment on another experimental manifestation
of the exchange self-energy. In principle, the Fermi velocity is
renormalized by this interaction \cite{hwang-prl99, dassarma-arXiv1203} 
and there has been some experimental hint of this in graphene in both
zero magnetic field \cite{elias-natphys7} and high magnetic fields.
\cite{jiang-prl98, li-g-prl102, zhou-natphys2}
The renormalized quasiparticle velocity can be written as
\begin{equation*}
	v_{F,\mathrm{int}} = \frac{1}{\hbar} \frac{d\mu_{\mathrm{HF}}}{d k}
	\sim \left(1 + r_s \log \frac{k_c}{k_F} \right)
\end{equation*}
with $r_s=e^2/(\hbar \kappa v_F)$, and thus the interaction-induced
correction will be proportional to $\log( k_c/k_F )$. 
Of course, the Fermi velocity measured in experiment includes this
renormalization so it is not directly measurable, but since the
self-energy is a nonmonotonic function of the wave vector, its effect
may be seen by a density-dependent deviation of the Fermi velocity
from a constant value which represents the combination of the SP
and mean interaction contributions.
Such a nonlinear reconstruction of the 2D TI Fermi surface at low
carrier density will be a direct manifestation of its Dirac spectrum and
the associated ultraviolet renormalization effect familiar in
quantum electrodynamics.

%\section{Comparisons with experiment \label{sec:expcomp}}

We can also extract \dmudn from ARPES measurements of the band
structure. Digitizing curves from Zhu \etal\cite{zhu-prl107} 
gives the band structure of \bise in the $K\Gamma K$ direction
from which we can numerically extract \dmudn for comparison to theory.
%However, this process is complicated by the fact that
%the curves on either side of the $\Gamma$ point are not
%perfectly symmetric, it is also known that there is a hexagonal
%distortion to the Fermi surface, \cite{chen-science325}
%and ARPES data contains some noise. 
To combat difficulties caused by imperfections in the experimental data,
we apply a Gaussian smoothing convolution with a width $\sigma =
0.08\mathrm{nm}^{-1}$ to both branches shown in the $K\Gamma K$ plot and
take the average of the two to arrive at the approximate band structure. 
%Then we associate the Fermi wave vector $k_F$ with this averaged curve
%for each value of energy and from this compute the density as
%$n=k_F^2/(4\pi)$ which assumes a circular Fermi surface.
The density is then computed and a numerical derivative of the chemical
potential with respect to the density may then be taken. The results of
this procedure are shown in Fig. \ref{fig:kappa-exp}(b) along with a
least-squares best fit of the quadratic correction dispersion using
\mast and $v_F$ as fitting parameters from the conduction band data. 
This fitting procedure yielded $\mast = 0.417m_e$ and $v_F = 1.00\times
10^6\mathrm{ms}^{-1}$, both of which are larger than the current common
estimates for these parameters in \bise. 
The dashed line corresponds to the predicted \dmudn for a purely
quadratic band. 
%which is constant at a value of $\dmudn= \frac{2\pi
%\hbar^2}{\mast} \approx 1.15\mathrm{eV}\mathrm{nm}^2$. 
This indicates that the contribution from the quadratic part of the band
is dominated by the linearity of the band structure over the whole
experimentally pertinent range of carrier density, clearly establishing
that the Dirac spectrum is dominating the physics in these systems.

%\section{Discussion and conclusions \label{sec:discussion}}

One likely challenge to using the compressibility to investigate the
topological surface states is the observed existence of substantial
disorder-induced inhomogeneity in the charge landscape of the surface.
\cite{beidenkopf-natphys7} Surface probes such as scanning tunneling
microscopy have shown that the screening
of an external disorder potential created by charged impurities in the
lattice of \bite and \bise lead to the formation of ``puddles'' of
electron and holes with spatial extent of the order of $10\mathrm{nm}$
and associated fluctuations in the local chemical potential of
approximately $10\mathrm{meV}$. This fluctuation has previously been
investigated in the context of charge transport,\cite{adam-prb85} but
in the case of our proposed SET measurements, it should be pointed out
that the area of the sample which influences the tip has a
radius\cite{yoo-science276, yacoby-ssc111} of the order of
$100\mathrm{nm}$, which is much larger than one puddle. 
Therefore, the SET will experience some averaged field corresponding to
the inhomogeneous landscape.
%We have previously examined a similar problem in the context of mono-
%and bilayer graphene \cite{abergel-prb83, abergel-prb84, abergel-prb86}
%and shown that for gapless systems, a phenomenological averaging method
%compares very well to SET measurements. Since the screening properties
%of the linear bands at low densities should be rather similar, we are
%confident that such approaches are valid in the context of \bite and
%\bise also. 
%In this case, the effect of the averaging will be to curtail
%the $1/\sqrt{n}$ divergence as $n\to 0$. At higher density, the
%averaging has very little effect.
The effect of the inhomogeneity is that even when the global average of
the density is zero, the local density is always finite and may be
positive or negative. Therefore, in the inhomogeneous case, the
local density never approaches zero and, hence, the $1/\sqrt{n}$
divergence is curtailed. At high density, the slow change of \dmudn with
density implies that the inhomogeneity will have little effect on the
measured compressibility. Therefore, the extraction of band parameters
from SET data is likely to be most accurate at higher density.

In this Rapid Communication, we have proposed that a capacitive surface
probe such as SET microscopy will reveal the properties of the
topological surface states of crystals such as \bite and \bise via
measurement of \dmudn. We have analytically
computed the SP contribution to \dmudn for model band
structures and described the role that electron-electron
interactions and charge inhomogeneity in the crystal will play. In
summary, the high dielectric constant implies that the quantitative
effect of interactions is negligible (although, at low density, there is
some hope of detecting a renormalization of the Fermi velocity due to
electron-electron interactions) and the inhomogeneity can be accounted
for via a straightforward phenomenological averaging technique which
will result in the low-density divergence being cut off.
We have established that compressibility measurements, when compared
with theory, could provide valuable information about both
SP and many-body properties of 2D TI surface states.

%\begin{acknowledgments}
We thank the U.S. ONR for support.
%\end{acknowledgments}

\bibliography{bibtex-sorted}

\end{document}